\documentclass[aps,prb,twocolumn]{revtex4}
\usepackage{amssymb,amsbsy,graphicx,times,subfigure,marvosym,color,bm,hyperref}
\usepackage[latin1]{inputenc}
\usepackage{sidecap}

\begin{document}

\title{Time reversal invariant realization of the Weyl semimetal phase}

\author{G\'{a}bor B. Hal\'{a}sz$^{1,2}$}
\author{Leon Balents$^1$}

\address{$^1$Kavli Institute for Theoretical Physics, University of California, Santa Barbara, CA 93106,
USA \\ $^2$Trinity College, University of Cambridge, Trinity Street,
Cambridge CB2 1TQ, UK}

%%%%%%%%%%%%%%%%%%%%%%%%%%%%%%%%%%%%%%%%%%%%%%%%

\begin{abstract}

We propose a realization of the Weyl semimetal phase that is
invariant under time reversal and occurs due to broken inversion
symmetry. We consider both a simple superlattice model and a more
realistic tight-binding model describing an experimentally
reasonable HgTe/CdTe multilayer structure. The two models have the
same underlying symmetry, therefore their low-energy features are
equivalent. We find a Weyl semimetal phase between the normal
insulator and the topological insulator phases that exists for a
finite range of the system parameters and exhibits a finite number
of Weyl points with robust band touching at the Fermi level. This
phase is experimentally characterized by a strong conductivity
anisotropy and topological surface states. The principal
conductivities change in a complementary fashion as the system
parameters are varied, and the surface states only exist in a region
of momentum space that is determined by the positions of the Weyl
points.

\end{abstract}

%%%%%%%%%%%%%%%%%%%%%%%%%%%%%%%%%%%%%%%%%%%%%%%%

\maketitle

%%%%%%%%%%%%%%%%%%%%%%%%%%%%%%%%%%%%%%%%%%%%%%%%

\section{Introduction} \label{sec-intro}

In the last decades, topological phases of matter have been in the
focus of intense theoretical and experimental study: for a review,
see Ref. \onlinecite{Hasan-Kane} and references therein. The order
exhibited by these phases is not associated with spontaneous
symmetry breaking, and it can be described by topological invariants
that are insensitive to smooth changes in the system
parameters.\cite{Thouless} As a generic feature, these phases also
have topologically protected edge states.

The field of topological phases was revolutionized by the discovery
of two-dimensional (2D) topological insulators,\cite{Topo-2D,
Bernevig} and the subsequent generalization to three-dimensional
(3D) topological insulators.\cite{Topo-3D} These materials exhibit a
bulk energy gap between the valence and the conduction bands,
similarly to normal insulators. On the other hand, they have gapless
surface states that are topologically protected, therefore
conduction is possible on the surface. Since topological insulators
arise due to strong spin-orbit coupling, their prevalence is larger
within materials consisting of heavier elements.\cite{Topo-exp} They
find potential applications in the areas of spintronics and quantum
computation.

It is a recent development that topologically protected surface
states can also be achieved in materials without a bulk energy gap:
these are the Weyl semimetals.\cite{Weyl, Burkov-Balents, Murakami}
They have band touching between the conduction and the valence bands
at the Fermi level. The band touching points are called Weyl points
because the dispersion relation around them is linear and hence the
excitations are equivalent to Weyl fermions. Weyl points can have
positive or negative helicities, and they always appear in pairs. To
achieve robust band touching that can not be removed by an
infinitesimal perturbation, Weyl points of opposite helicities must
be separated in momentum space.\cite{Burkov-Balents} This requires
breaking either the time reversal or the inversion symmetry of the
system.\cite{Herring}

Recent papers on Weyl semimetals have predominantly studied the case
with broken time reversal symmetry.\cite{Weyl, Burkov-Balents} One
notable exception is Ref. \onlinecite{Murakami} where time reversal
symmetry remains intact and inversion symmetry is broken. It was
argued that a gapless phase appears in three dimensions between the
normal insulator (NI) and the topological insulator (TI) phases. In
this paper, we are also interested in the time reversal invariant
case, but address specifically how this phase, which is in fact the
Weyl semimetal, may be designed in a NI/TI superlattice.  We propose
two models: a simple superlattice model adapted from Ref.
\onlinecite{Burkov-Balents} and a more realistic tight-binding model
describing a HgTe/CdTe multilayer structure. The former model is
presented in Section \ref{sec-super} and the latter one is presented
in Section \ref{sec-tight}. The most prominent experimental features
are discussed in Section \ref{sec-phys}, while the overall
conclusions of the paper are summarized in Section \ref{sec-summ}.

\section{Superlattice model} \label{sec-super}

\subsection{General description} \label{ssec-gen}

The model considered in this section is based on the multilayer
structure in Ref. \onlinecite{Burkov-Balents}: a periodic
superlattice of NI and TI layers grown in the $z$ direction. It is a
simplified tight-binding model where we only take the surface states
located at the NI/TI interfaces into account. These states are
labeled by the unit cell index and the parallel 2D momentum
$\mathbf{k} = (k_x, k_y)$.

It is known that a realization of the Weyl semimetal phase requires
breaking either the time reversal or the inversion symmetry of the
system. Since we intend to keep the time reversal symmetry intact,
the inversion symmetry must be broken. To achieve that, we introduce
a finite voltage $V$ between the top and the bottom NI/TI interfaces
in each unit cell. The Hamiltonian of the multilayer structure is
then
\begin{eqnarray}
H &=& \sum_{\mathbf{k}} \sum_{i,j} \bigg[ v_F \tau^z (\sigma^x k_y -
\sigma^y k_x) \delta_{i,j} + V \tau^z \delta_{i,j} \nonumber \\
&+& \Delta_T \, \tau^x \delta_{i,j} + \Delta_N \sum_{\pm} \tau^{\pm}
\delta_{i, j \pm 1} \bigg] c^{\dag}_{i, \mathbf{k}} c_{j,
\mathbf{k}}, \label{eq-H-1}
\end{eqnarray}
where the Pauli matrices $\vec{\sigma} = (\sigma^x, \sigma^y,
\sigma^z)$ act on the real spin degree of freedom and the Pauli
matrices $\vec{\tau} = (\tau^x, \tau^y, \tau^z)$ act on the
top/bottom surface pseudospin degree of freedom. The first term
describes the NI/TI surface states with isotropic Fermi velocity
$v_F$, the second term represents the inversion-breaking voltage,
and the remaining terms describe hopping between neighboring
interfaces. The hopping amplitude through a TI layer is $\Delta_T$
and that through a NI layer is $\Delta_N$. In general, both
$\Delta_T$ and $\Delta_N$ can be functions of the parallel momentum
$\mathbf{k}$, and the symmetries of these functions determine the
symmetry of the system.

The Hamiltonian in Eq. (\ref{eq-H-1}) can be solved by exploiting
the translational symmetry in the $z$ direction, and introducing the
corresponding 3D momentum $\vec{k} = (k_x, k_y, k_z)$. By doing so,
we find that the band dispersion relation is
\begin{equation}
E_{\pm}^2 (\vec{k}) = \Delta^2 (k_z) + \big[ V \pm v_F |\mathbf{k}|
\big]^2, \label{eq-disp-11}
\end{equation}
where $\Delta (k_z) = \sqrt{\Delta_T^2 + \Delta_N^2 + 2 \Delta_T
\Delta_N \cos (k_z d)}$ and $d$ is the periodicity of the
superlattice. The four bands are non-degenerate when $\mathbf{k}
\neq 0$, and band touching between the two middle bands takes place
when $E_{-} = 0$. If we assume without loss of generality that
$\Delta_T$ and $\Delta_N$ are both positive, this happens when $k_z
d = \pi$, $\Delta_T = \Delta_N$, and $V = v_F |\mathbf{k}|$.

\subsection{The Weyl semimetal phase} \label{ssec-weyl-1}

If $\Delta_T$ and $\Delta_N$ are independent of $\mathbf{k}$, the
band touching occurs along a circle of radius $V / v_F$ in the $k_z
= \pi / d$ plane. It marks the transition between the NI and the TI
phases of the material at $\Delta_T = \Delta_N$. We can argue on
physical grounds that $\Delta_T > \Delta_N$ (thin TI layers and
thick NI layers) corresponds to the NI phase, while $\Delta_T <
\Delta_N$ (thick TI layers and thin NI layers) corresponds to the TI
phase.

However, this band touching is not robust because it requires the
fine-tuning of the condition $\Delta_T = \Delta_N$. To achieve
robust band touching, we need to make the hopping amplitudes depend
on the momentum $\mathbf{k}$:
\begin{equation}
\Delta_{T,N} = \Delta_{T,N}^{(0)} + \Delta_{T,N}^{(1)} (\mathbf{k}).
\label{eq-delta-0}
\end{equation}
Furthermore, we can not keep the continuous rotational symmetry
around the $z$ axis because then $\Delta_T$ and $\Delta_N$ are still
constants at $|\mathbf{k}| = V / v_F$, the only region where band
touching is possible. On the other hand, the continuous rotational
symmetry is broken in real crystals as well, and one is only left
with a discrete rotational symmetry. In the following, we
demonstrate robust band touching in the reasonable cases of the
four-fold and two-fold rotational symmetries.

\subsection{Four-fold rotational symmetry} \label{ssec-four}

In the first case, we assume a four-fold rotational symmetry around
the $z$ axis and four planes of reflection symmetry: the $\{x,z\}$
plane, the $\{y,z\}$ plane, and the two planes halfway in between.
These are the natural symmetries of many real materials with
tetragonal crystal structures. By neglecting any contributions
depending on $|\mathbf{k}|$ only, the lowest order term having all
the above symmetries and time reversal symmetry is $\propto (k_x^4 +
k_y^4)$. The $\mathbf{k}$ dependent parts of the hopping amplitudes
are then
\begin{equation}
\Delta_{T,N}^{(1)} (\mathbf{k}) = \delta_{T,N} |\mathbf{k}|^4 \left(
\cos^4 \theta + \sin^4 \theta \right), \label{eq-delta-1}
\end{equation}
where the polar coordinates $k_x = |\mathbf{k}| \cos \theta$ and
$k_y = |\mathbf{k}| \sin \theta$ are introduced. The difference
$\Delta_T - \Delta_N$ depends on the angle $\theta$ at $|\mathbf{k}|
= V / v_F$, therefore band touching with $\Delta_T = \Delta_N$ only
occurs at specific points of the circle. The band touching also
becomes robust because the parameters $\Delta_{T,N}^{(0)}$ and
$\delta_{T,N}$ do not require fine-tuning: an infinitesimal change
in any of them only gives an infinitesimal change in $\theta$,
displacing the band touching points along the circle.

\begin{figure}[h]
\centering
\includegraphics[width=7.8cm]{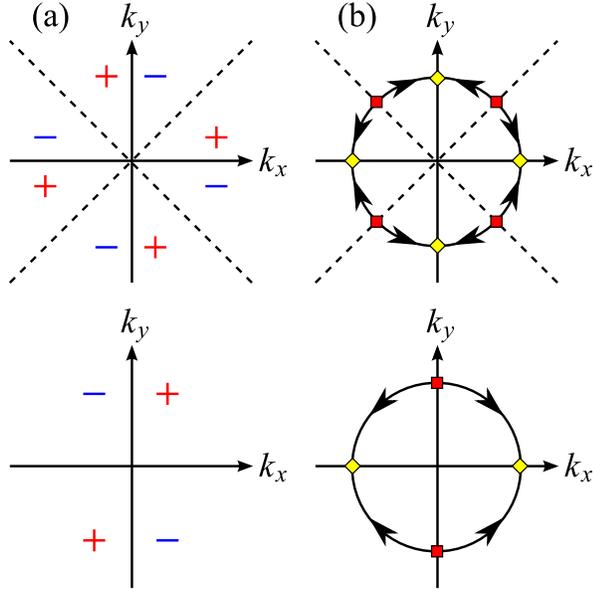}
\caption{(Color online) Illustration of the Weyl semimetal phase in
the cases of four-fold (top) and two-fold (bottom) rotational
symmetries. We set $k_z = \pi / d$ in all subfigures. (a)
Arrangement of the Weyl points with positive (red plus) and negative
(blue minus) helicities. (b) Trajectories of the Weyl points as the
transition from the NI phase to the TI phase takes place. The Weyl
points first appear at the red squares and finally disappear at the
yellow diamonds. \label{fig-1}}
\end{figure}

Contrary to the case with continuous rotational symmetry, now there
is a Weyl semimetal phase between the NI and the TI phases that
exists for a finite range of the parameter values. This phase
features a finite number of Weyl points at which band touching
between the two middle bands occurs. To be more precise, the
solution of the band touching equation $\Delta_T = \Delta_N$ for the
angle $\theta$ is
\begin{equation}
\cos (4 \, \theta) = \frac{4 \big( \Delta_N^{(0)} - \Delta_T^{(0)}
\big)} {|\mathbf{k}|^4 (\delta_T - \delta_N)} - 3.
\label{eq-theta-1}
\end{equation}
This expression gives 8 Weyl points which are related to each other
by the symmetry transformations of the system. Since $|\cos (4 \,
\theta) \, | \leq 1$, the condition for the Weyl semimetal phase
becomes $1/2 < \big( \Delta_N^{(0)} - \Delta_T^{(0)} \big) / \left[
|\mathbf{k}|^4 (\delta_T - \delta_N) \right] < 1$, where
$|\mathbf{k}| = V / v_F$ as always in this subsection. Let us assume
without loss of generality that $\delta_T > \delta_N$, and imagine
decreasing $\Delta_T^{(0)}$ gradually while keeping the other
parameters constant. This corresponds to a transition from the NI
phase to the TI phase. The Weyl points then first appear at the
lines $k_x = \pm k_y$, move along the circle of radius $V / v_F$,
and finally disappear at the lines $k_x = 0$ and $k_y = 0$. For an
illustration of this, see the top half of Fig. \ref{fig-1}.

It can be verified that the band touching points occurring in this
scenario are indeed proper Weyl points around which the band
dispersion is linear in all directions. To obtain physically
transparent results, we assume that $\delta_T - \delta_N$ is
sufficiently small so that $\Delta_T - \Delta_N$ is almost
independent of $\theta$. This difference is then only important
along circles of constant $k_z$ and $|\mathbf{k}|$ where there would
be no difference otherwise. Consequently, the principal directions
are the axial ($z$), the radial ($r$), and the tangential ($t$)
directions, and an expansion of $E_{-}$ around a band touching point
reads
\begin{equation}
E_{-}^2 (\vec{k}) = v_z^2 \delta k_z^2 + v_r^2 \delta k_r^2 + v_t^2
\delta k_t^2, \label{eq-disp-12}
\end{equation}
where the effective Fermi velocities corresponding to the principal
directions are $v_z = d \sqrt{\Delta_T \Delta_N} = d \Delta_T$, $v_r
= v_F$, and $v_t = |\mathbf{k}|^3 (\delta_T - \delta_N) |\sin (4 \,
\theta)|$. The expression in Eq. (\ref{eq-disp-12}) indeed gives a
linear band dispersion in all directions. We can now establish that
the approximation of small $\delta_T - \delta_N$ requires $v_t \ll
v_r$, i.e. $V^3 (\delta_T - \delta_N) \ll v_F^4$. This is satisfied
in the reasonable case when the inversion-breaking voltage $V$ and
the coefficients $\delta_{T,N}$ are small.

We note that the Weyl points related to each other by rotations have
identical helicities, while those related to each other by
reflections have opposite helicities: this implies that there are 4
Weyl points of each helicity. If we pair up all the Weyl points into
pairs of opposite helicities, the sum of the resulting separation
vectors is zero. This property follows from the general notion of
time reversal symmetry which also implies that the anomalous Hall
conductivity vanishes.

\subsection{Two-fold rotational symmetry} \label{ssec-two}

In the second case, we have a two-fold rotational symmetry around
the $z$ axis and two planes of reflection symmetry: the $\{x,z\}$
plane and the $\{y,z\}$ plane. This case is particularly important
for us because the tight-binding model described in Section
\ref{sec-tight} has the same symmetries. The lowest order terms
obeying all these symmetries and time reversal symmetry are $\propto
(k_x^2 + k_y^2)$ and $\propto (k_x^2 - k_y^2)$. However, the former
one only depends on $|\mathbf{k}|$, and hence it would not break the
continuous rotational symmetry on its own. For the sake of
simplicity, we consider the special case of
\begin{equation}
\Delta_{T,N}^{(1)} (\mathbf{k}) = \delta_{T,N} \, k_x^2 =
\delta_{T,N} |\mathbf{k}|^2 \cos^2 \theta, \label{eq-delta-2}
\end{equation}
and obtain that the solution of $\Delta_T = \Delta_N$ is
\begin{equation}
\cos (2 \, \theta) = \frac{2 \big( \Delta_N^{(0)} - \Delta_T^{(0)}
\big)} {|\mathbf{k}|^2 (\delta_T - \delta_N)} - 1.
\label{eq-theta-2}
\end{equation}
Now there are 4 Weyl points in the Weyl semimetal phase that occurs
when $0 < \big( \Delta_N^{(0)} - \Delta_T^{(0)} \big) / \left[
|\mathbf{k}|^2 (\delta_T - \delta_N) \right] < 1$. If there is a
transition from the NI phase to the TI phase due to a gradual
decrease in $\Delta_T^{(0)}$, the Weyl points first appear at the
$k_x = 0$ line, move along the circle of radius $V / v_F$, and
finally disappear at the $k_y = 0$ line. For an illustration of
this, see the bottom half of Fig. \ref{fig-1}.

If $\delta_T - \delta_N$ is sufficiently small, the principal
directions around the Weyl points are the axial, the radial, and the
tangential directions. Eq. (\ref{eq-disp-12}) is therefore valid in
this case as well, and the effective Fermi velocities in the
principal directions are given by $v_z = d \sqrt{\Delta_T \Delta_N}
= d \Delta_T$, $v_r = v_F$, and $v_t = |\mathbf{k}| (\delta_T -
\delta_N) |\sin (2 \, \theta)|$. The approximation of small
$\delta_T - \delta_N$ holds when $v_t \ll v_r$, i.e. when $V
(\delta_T - \delta_N) \ll v_F^2$.

\section{Realistic tight-binding model} \label{sec-tight}

\subsection{Formulation of the model} \label{ssec-form}

In this section, we consider a periodic multilayer structure of
strained HgTe and CdTe layers which are grown on top of each other
in the $z$ direction. This model is in fact a concrete realization
of the superlattice structure described in Section \ref{sec-super}
because CdTe is a NI and HgTe becomes a TI under strain.\cite{Dai}
The band structures of these materials are well known, and can be
reproduced with high accuracy from realistic tight-binding models.
Here we adapt the ten-band tight-binding model described in Ref.
\onlinecite{Kobayashi} which assumes two $s$ orbitals ($s$, $s^*$)
and three $p$ orbitals ($p_x$, $p_y$, $p_z$) on each atom.

Both HgTe and CdTe have zinc-blende structures: the anions (Te) form
a face-centered cubic lattice, and the cations (Hg, Cd) are located
at the positions $\frac{1}{4} [1, 1, 1]$. This implies that each
anion (cation) is tetrahedrally coordinated by 4 nearest neighbor
cations (anions). We assume that the thicknesses of the HgTe and
CdTe layers are $N_1$ and $N_2$ as measured in units of the cubic
lattice parameter $a$. When cutting through the structure along the
$z$ direction, one finds subsequent layers of one atomic thickness
consisting of only anions and only cations, respectively. The
anionic layers all contain Te, while there are $2N_1$ cationic
layers containing Hg and $2N_2$ cationic layers containing Cd in
each superlattice period.

The atomic orbitals are labeled according to $| \vec{R}, u, t,
\sigma \rangle$, where $\vec{R}$ is the position of the atomic site,
$u = \{ \textrm{Te}, \textrm{Hg}, \textrm{Cd} \}$ is the type of the
atom, $t = \{ s, s^*, p_x, p_y, p_z \}$ is the type of the orbital,
and $\sigma = \{ \uparrow, \downarrow \}$ is the spin quantum
number. The Hamiltonian of the system can be written as
\begin{equation}
H = H_0 + H_I + H_{\mathrm{so}}, \label{eq-H-21}
\end{equation}
where the first term contains the bare energies of the atomic
orbitals, the second term describes the interaction (hopping)
between them, and the third term represents spin-orbit coupling.
These terms are
\begin{eqnarray}
H_0 &=& \sum_{\vec{R}, t, \sigma} | \vec{R}, u (\vec{R}), t, \sigma
\rangle E_{u (\vec{R}), t} \langle \vec{R}, u (\vec{R}), t, \sigma
|, \label{eq-H-22} \\
H_I &=& \sum_{\vec{R}, t, \sigma} \sum_{\vec{R}', t'} | \vec{R}, u
(\vec{R}), t, \sigma \rangle V_{u (\vec{R}), u (\vec{R}'), t, t'}
\langle \vec{R}', u (\vec{R}'), t', \sigma |, \nonumber \\
H_{\mathrm{so}} &=& \sum_{\vec{R}, t, t' \sigma, \sigma'} | \vec{R},
u (\vec{R}), t, \sigma \rangle 2\lambda_{u (\vec{R})} \vec{L} \cdot
\vec{\sigma} \langle \vec{R}, u (\vec{R}), t', \sigma' |, \nonumber
\end{eqnarray}
where the second sum in $H_I$ goes over all 4 nearest neighbors
$\vec{R}'$ of the atomic site $\vec{R}$, and $u (\vec{R})$ denotes
the type of the atom at the position $\vec{R}$.

The hopping amplitudes $V_{u, u', t, t'}$ between different types of
$p$ orbitals are related to each other by the geometry of the
crystal structure. In particular, they are affected by the uniaxial
strain $\epsilon$ which is defined as the relative elongation of the
lattice constant in the $z$ direction with respect to those in the
$x$ and $y$ directions. This strain occurs because the subsequent
layers of HgTe and CdTe are grown on top of each other, and there is
a slight lattice constant mismatch.\cite{Schulman} Since the lattice
constant of CdTe is approximately 0.3\% larger than that of HgTe,
and both materials have a Poisson's ratio $\approx 0.5$, we assume
that the relationship between the strains in them is $\epsilon
(\textrm{CdTe}) = \epsilon (\textrm{HgTe}) + 0.009$. Even if a small
strain does not change the distance between neighboring atoms, the
direction vector connecting them changes, leading to a different
overlap between any two orbitals if at least one of them is a $p$
orbital. Using simple geometry, the different hopping amplitudes
involving $p$ orbitals are then expressed as
\begin{eqnarray}
V_{s, p_x} &=& V_{s, p_y} = \frac{1}{\sqrt{3}} V_{s, p, \sigma}
\left( 1 - \frac{\epsilon} {3} \right), \label{eq-V} \\
V_{s, p_z} &=& \frac{1}{\sqrt{3}} V_{s, p, \sigma}
\left( 1 + \frac{2\epsilon} {3} \right), \nonumber \\
V_{p_x, p_x} &=& V_{p_y, p_y} = \frac{1}{3} \left[ V_{p, p, \sigma}
\left( 1 - \frac{2\epsilon} {3} \right) + 2V_{p, p, \pi}
\left( 1 + \frac{\epsilon} {3} \right) \right], \nonumber \\
V_{p_z, p_z} &=& \frac{1}{3} \left[ V_{p, p, \sigma} \left( 1 +
\frac{4\epsilon} {3} \right) + 2V_{p, p, \pi}
\left( 1 - \frac{2\epsilon} {3} \right) \right], \nonumber \\
V_{p_x, p_y} &=& \frac{1}{3} \left( V_{p, p, \sigma} - V_{p, p, \pi}
\right) \left( 1 - \frac{2\epsilon} {3} \right), \nonumber \\
V_{p_x, p_z} &=& V_{p_y, p_z} = \frac{1}{3} \left( V_{p, p, \sigma}
- V_{p, p, \pi} \right) \left( 1 + \frac{\epsilon} {3} \right).
\nonumber
\end{eqnarray}
Note that the subscripts $u$ and $u'$ are suppressed for the sake of
compactness, the label $s$ can stand for both $s$ and $s^*$, and all
terms are expanded up to first order in $\epsilon$.

\begin{table}[h]
\begin{tabular*}{0.48\textwidth}{@{\extracolsep{\fill}}c c c c}
\hline \hline
$u$            &   Te        &   Hg        &   Cd        \\
\hline
$E_{u, s}$     &   $-$9.75   &   $-$1.40   &   $-$1.42   \\
$E_{u, p}$     &   0.12      &   4.30      &   3.48      \\
$E_{u, s^*}$   &   6.08      &   6.50      &   6.67      \\
$\lambda_u$    &   0.333     &   0.286     &   0.013     \\
\hline  \hline
\end{tabular*}
\caption{Bare energies $E_{u, t}$ and spin-orbit coupling strengths
$\lambda_u$ for different atom types (all numbers are in eV units).
\label{table-1}}
\end{table}

The concrete tight-binding parameters are based on those in Ref.
\onlinecite{Kobayashi}, but they are normalized according to a
consistent procedure. The bare energies of all orbitals in the CdTe
model are first shifted such that the Te orbitals have the same
average energy in HgTe and CdTe. This corresponds to matching the
arbitrary zero energy levels of the independent HgTe and CdTe
tight-binding models. The bare energies of the respective Te
orbitals are then obtained by averaging those in the normalized HgTe
and CdTe models which are already close to each other at this point.
The spin-orbit coupling strength $\lambda_{\mathrm{Te}}$ is averaged
in the same way, and the final values of the tight-binding
parameters are presented in Tables \ref{table-1} and \ref{table-2}.

\begin{table}[h]
\begin{tabular*}{0.48\textwidth}{@{\extracolsep{\fill}}c c c}
\hline \hline
$u / u'$                      &   Te/Hg      &   Te/Cd      \\
\hline
$V_{u, u', s, s}$             &   $-$0.817   &   $-$1.195   \\
$V_{u, u', s, p, \sigma}$     &   1.044      &   0.753      \\
$V_{u, u', p, s, \sigma}$     &   $-$1.404   &   $-$2.064   \\
$V_{u, u', s^*, p, \sigma}$   &   1.524      &   0.844      \\
$V_{u, u', p, s^*, \sigma}$   &   $-$0.140   &   $-$1.147   \\
$V_{u, u', p, p, \sigma}$     &   2.180      &   2.651      \\
$V_{u, u', p, p, \pi}$        &   $-$0.549   &   $-$0.442   \\
\hline  \hline
\end{tabular*}
\caption{Hopping amplitudes between different atom and orbital types
(all numbers are in eV units). \label{table-2}}
\end{table}

Table \ref{table-1} shows that the normalized bare energies of the
Hg orbitals are on average larger than those of their Cd
counterparts. The HgTe layers are therefore more positively charged
than the CdTe layers, resulting in a potential difference that
lowers the orbital energies in the HgTe layers. This effect is taken
into account by introducing a periodic potential $U (\vec{r})$ which
is added to all bare energies at position $\vec{r}$. The potential
depends on $z$ only, and we write it in the form
\begin{equation}
U (z) = -U_0 \cos \left[ \frac{2\pi} {d} \left( z - \frac{d_1}{2} +
\delta_0 \right) \right], \label{eq-pot}
\end{equation}
where $d_{1,2} = a N_{1,2}$ are the thicknesses of the HgTe and CdTe
layers, and $d = d_1 + d_2$ is the periodicity of the superlattice.
In the symmetric case when $\delta_0 = 0$, the potential function $U
(z)$ reaches its minimum in the middle of the HgTe layer and its
maximum in the middle of the CdTe layer. However, we assume a
certain asymmetry in $U (z)$ which is characterized by the
displacement $\delta_0$ of these extrema. When the multilayer is
grown under reasonable experimental conditions, such an asymmetry is
inadvertently present due to the specific growth direction. For
example, it is possible that the HgTe and CdTe materials are more
likely to form an alloy at one of their interfaces. Since the
resulting asymmetry is probably small, we take $0 < |\delta_0| < a$
in the rest of this section. Also, by comparing the bare energies in
Table \ref{table-1} we estimate that the amplitude of the potential
is $U_0 \sim 0.1$ eV.

\subsection{Linear four-band approximation} \label{ssec-lin}

The Hamiltonian presented in Eqs. (\ref{eq-H-21}) and
(\ref{eq-H-22}) can be solved by exploiting translational invariance
and introducing the corresponding momentum $\vec{k} = (k_x, k_y,
k_z)$. On the other hand, the large periodicity in the $z$ direction
means that the Hamiltonian is represented by a large $M \times M$
matrix where $M = 40 (N_1 + N_2)$. It can therefore only be solved
numerically, and for relatively small layer thicknesses $N_{1,2}$.

However, despite the complexity in this model, some of its
properties can be deduced by referring to symmetry only. In the case
of $U_0 \neq 0$ and $\delta_0 \neq 0$, the basic symmetries of the
system are time reversal symmetry ($\mathrm{T}$) and the reflection
symmetries ($\mathrm{R}_{1,2}$) across the $\{y',z\}$ and the
$\{x',z\}$ planes. The natural coordinates $x' = (x + y) / \sqrt{2}$
and $y' = (x - y) / \sqrt{2}$ are introduced to make these
symmetries more explicit. Note that the reflection symmetries
$\mathrm{R}_{1,2}$ also lead to a two-fold rotational symmetry
($\mathrm{S}$) around the $z$ axis. In terms of symmetry, the
tight-binding model studied in this section is equivalent to the
superlattice model in Section \ref{ssec-two}.

The presence of both $\mathrm{T}$ and $\mathrm{S}$ symmetries puts
two crucial restrictions on the band structure. First, the states on
the $k_{x'} = k_{y'} = 0$ line are two-fold degenerate. Since the
number of occupied bands is always even, the highest occupied band
(HOB) and the lowest unoccupied band (LUB) have different energies
on this line with the Fermi level lying between them. Second, the
band structure is invariant under the reflection $k_z
\leftrightarrow -k_z$. As shown in Section \ref{ssec-rob}, this
implies that robust band touching between the HOB and the LUB is
only possible in the $k_z = 0$ and the $k_z = \pi / d$ planes.

Numerical investigation of the model indicates that band touching
between the HOB and the LUB always occurs close to the $k_{x'} =
k_{y'} = 0$ line. In perspective of this and the symmetry
considerations above, we introduce simplified four-band models
around the two special points at $\vec{k} = (0, 0, 0)$ and $\vec{k}
= (0, 0, \pi / d)$. We only keep the nearest two bands on each side
of the Fermi level, and assume that they are linear in the relative
momentum $\delta \vec{k} = (k_{x'}, k_{y'}, \delta k_z)$ with
respect to the corresponding special point. More formally, we
project the Hamiltonian $H$ onto a subspace spanned by four basis
states: the appropriate eigenvectors of the full model at $\delta
\vec{k} = 0$. The reduced Hamiltonian $\mathcal{H}$ is then a $4
\times 4$ diagonal matrix at $\delta \vec{k} = 0$, and the linearity
of the band dispersion is achieved by additional terms that are
linear in $\delta \vec{k}$.

At the special point, the reduced Hamiltonian can be written as
$\mathcal{H}_0 = E^{(1)} \tau^{(1)} + E^{(2)} \tau^{(2)}$ where
$\tau^{(1,2)} = (1 \pm \tau^{z}) / 2$ and $E^{(1,2)}$ are the
energies of the LUB and the HOB at $\delta \vec{k} = 0$. The energy
levels are pairwise degenerate, and this degeneracy is split by a
finite $k_{x'}$ or $k_{y'}$ but not by a finite $\delta k_z$. To
represent this splitting, we need to add coupling terms between
states corresponding to the same energy at $\delta \vec{k} = 0$. In
the most general case, the additional terms in the Hamiltonian read
\begin{equation}
\mathcal{H}_S = \sum_{l=1}^2 \tau^{(l)} \left[ \alpha_x^{(l)} k_{x'}
\sigma^x + \alpha_y^{(l)} k_{y'} \sigma^y \right], \label{eq-H-31}
\end{equation}
where the coefficients $\alpha_{x,y}^{(1,2)}$ can be obtained from a
comparison with the full model. The choice of the Pauli matrices
$\sigma^{x,y}$ corresponds to defining the basis states within the
degenerate subspaces of $\mathcal{H}_0$ in a particular way.

\begin{table}[h]
\begin{tabular*}{0.48\textwidth}{@{\extracolsep{\fill}}c c c c c}
\hline \hline
                &   $\mathrm{T}$   &   $\mathrm{R}_1$   &   $\mathrm{R}_2$   &   $\mathrm{S}$   \\
\hline
$k_{x'}$       &   $-$            &   $-$              &   +                &   $-$            \\
$k_{y'}$       &   $-$            &   +                &   $-$              &   $-$            \\
$\delta k_z$   &   $-$            &   +                &   +                &   +              \\
$\sigma^x$     &   $-$            &   $-$              &   +                &   $-$            \\
$\sigma^y$     &   $-$            &   +                &   $-$              &   $-$            \\
$\sigma^z$     &   $-$            &   $-$              &   $-$              &   +              \\
$\tau^x$       &   +              &   $-$              &   $-$              &   +              \\
$\tau^y$       &   $-$            &   $-$              &   $-$              &   +              \\
$\tau^z$       &   +              &   +                &   +                &   +              \\
\hline  \hline
\end{tabular*}
\caption{Potential terms in the Hamiltonian and their parities under
the symmetry operations of the system: time reversal ($\mathrm{T}$),
reflection across the $\{y',z\}$ plane ($\mathrm{R}_1$), reflection
across the $\{x',z\}$ plane ($\mathrm{R}_2$), and two-fold rotation
around the $z$ axis ($\mathrm{S}$). \label{table-3}}
\end{table}

There is also coupling between states corresponding to different
energies at $\delta \vec{k} = 0$. Since the Hamiltonian
$\mathcal{H}$ must be invariant under all symmetry operations of the
system, there are only a small number of such coupling terms allowed
by symmetry. The parities of the possible terms under the symmetry
operations are summarized in Table \ref{table-3}. Note that the
parities of $\tau^z$ and $\sigma^{x,y}$ are determined by the
already established terms $\mathcal{H}_0$ and $\mathcal{H}_S$ which
must be even under all symmetry operations. Furthermore, the
relations between different Pauli matrices imply that we only need
to choose the parities of $\tau^x$ under $\mathrm{T}$ and
$\mathrm{R}_{1,2}$. The choice of these parities corresponds to
setting the relative complex phases of the basis states. Under the
choice presented in Table \ref{table-3}, the most general
contribution to the Hamiltonian takes the form
\begin{equation}
\mathcal{H}_D = \tau^{x} \left[ \beta_x k_{y'} \sigma^x + \beta_y
k_{x'} \sigma^y + \beta_z \delta k_z \sigma^z \right],
\label{eq-H-32}
\end{equation}
where the coefficients $\beta_{x,y,z}$ are again to be determined
from a comparison with the full model. The reduced Hamiltonian
finally reads $\mathcal{H} = \mathcal{H}_0 + \mathcal{H}_S +
\mathcal{H}_D$. It is a considerable simplification with respect to
$H$, and it only contains 9 parameters that need to be extracted
from the full model.

\subsection{Conditions for robust band touching} \label{ssec-rob}

Band touching between the HOB and the LUB occurs in the full model
when the two middle eigenvalues are equal in the simplified model.
It can be shown that for a $4 \times 4$ matrix of the form
$\mathcal{H}$, this is possible if and only if the direction of the
vector $\vec{B} = (\beta_x k_{y'}, \beta_y k_{x'}, \beta_z \delta
k_z)$ lies halfway between the directions of the vectors
$\vec{A}^{(1,2)} = (\alpha_x^{(1,2)} k_{x'}, \alpha_y^{(1,2)}
k_{y'}, 0)$. The two bands then cross each other as $|\delta
\vec{k}|$ is increased without changing the direction of $\delta
\vec{k}$, whereas anti-crossing happens otherwise. Since the above
condition requires the three vectors to lie in the same plane, the
third component of $\vec{B}$ has to vanish. Due to $\beta_z \neq 0$
in general, we find that robust band touching can only occur in the
$\delta k_z = 0$ plane.

Restricting our attention to this plane simplifies the problem
because $\vec{A}^{(1,2)}$ and $\vec{B}$ become 2D vectors. If we
change the ratio $k_{y'} / k_{x'}$ gradually from $0$ to $\infty$,
the ratios of the corresponding components in $\vec{A}^{(1,2)}$
change in the same direction, while those in $\vec{B}$ change in the
opposite direction between $0$ and $\pm \infty$. This means that
whether band touching happens at any $\delta \vec{k}$ is determined
entirely by the signs of the different parameters. Since we always
choose $\alpha_{x,y}^{(1,2)} > 0$, the condition becomes
straightforward: band touching occurs if and only if $\beta_x$ and
$\beta_y$ have the same sign.

Let us now consider the special case of the symmetric potential with
$\delta_0 = 0$. By repeating the symmetry considerations in Section
\ref{ssec-lin} and taking into account the additional four-fold
roto-reflection symmetry around the $z$ axis, we find that the
parameters from the full model are no longer independent because
$\alpha_x^{(1,2)} = \alpha_y^{(1,2)}$ and $\beta_x = -\beta_y$. This
shows that band touching can only occur in this scenario if at least
one of these parameters vanishes. However, the corresponding band
touching is not robust because it requires the fine-tuning of a
parameter. We conclude that robust band touching requires the
asymmetry characterized by $\delta_0 \neq 0$, and expect that it
becomes easier to observe as $U_0$ and $\delta_0$ increase.

\subsection{The Weyl semimetal phase} \label{ssec-weyl-2}

The detailed behavior of the system is determined by how the
coefficients from the full model depend on the external parameters.
Since this dependence is affected by the complexity of the full
model, its understanding requires a numerical treatment. In
perspective of this, we numerically investigate the phenomenon of
robust band touching in the function of the layer thicknesses
$N_{1,2}$, the strain $\epsilon_0 \equiv \epsilon (\textrm{HgTe})$
in the multilayer structure, the amplitude $U_0$ of the superlattice
potential, and the asymmetric displacement $\delta_0$.

\begin{figure}[b]
\centering
\includegraphics[width=8.9cm]{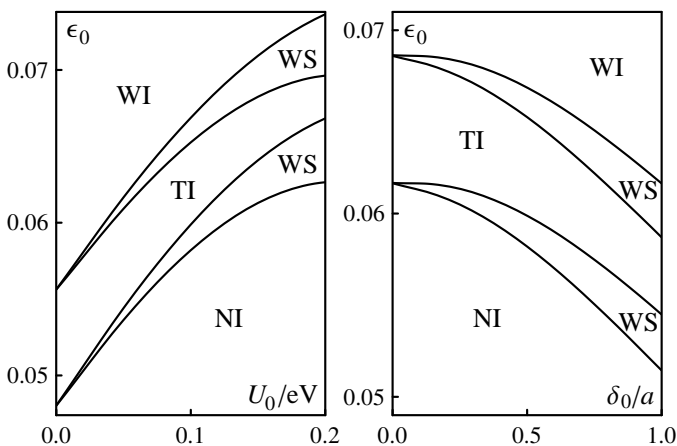}
\caption{Critical strains $\epsilon_0$ against $U_0$ at constant
$\delta_0 = a / 2$ (left) and against $\delta_0$ at constant $U_0 =
0.1$ eV (right). The phase boundaries separate four distinct phases:
the normal insulator (NI), the strong (3D) topological insulator
(TI), the weak (2D) topological insulator (WI), and the Weyl
semimetal (WS). The layer thicknesses are constant $N_1 = 3$ and
$N_2 = 4$ in both subfigures. \label{fig-2}}
\end{figure}

We first consider the dependence on the strain. If $U_0 \neq 0$ and
$\delta_0 \neq 0$, there are two ranges in $\epsilon_0$ close to
zero with band touching in the $k_z = 0$ and the $k_z = \pi / d$
planes, respectively. The corresponding band touching is robust
because it remains intact for an infinitesimal change in any of the
external parameters $\epsilon_0$, $U_0$, and $\delta_0$. The upper
and lower limits of the ranges are functions of $U_0$ and $\delta_0$
as illustrated in Fig. \ref{fig-2}, and we verify the expectation
from Section \ref{ssec-rob} that the ranges increase with both $U_0$
and $\delta_0$. For the reasonable values of $U_0 \sim 0.1$ eV and
$\delta_0 \sim a / 2$, the ranges are $\Delta \epsilon_0 \sim
0.002$.

Now we turn our attention to the layer thicknesses. Keeping the HgTe
thickness $N_1 = 3$ constant and varying the CdTe thickness $N_2$
between $4$ and $8$ reveals that an increase in $N_2$ decreases
$\Delta \epsilon_0$. This is intuitive because $\delta_0$ becomes
smaller in comparison to $d$. Keeping the CdTe thickness $N_2 = 4$
constant and varying the HgTe thickness $N_1$ between $3$ and $7$
shows that an increase in $N_1$ shifts the ranges in $\epsilon_0$
downwards. This means that the phases with robust band touching
appear at more negative strains.

To conclude that these phases are indeed Weyl semimetals, they need
to satisfy one more condition: the lack of band overlap. Even if
there is robust band touching between the HOB and the LUB, the band
structure becomes metallic if the highest overall energy of the HOB
is larger than the lowest overall energy of the LUB. It is an
empirical observation that the individual band structures of the
$k_z = 0$ and the $k_z = \pi / d$ planes are metallic when
$\epsilon_0$ is sufficiently negative. This occurs for all
$\epsilon_0 < 0$ in the limit of $U_0 \rightarrow 0$ or $\delta_0
\rightarrow 0$, while the critical $\epsilon_0$ becomes slightly
negative at larger values of $U_0$ and $\delta_0$. Furthermore, the
appropriate bands of the $k_z = 0$ and the $k_z = \pi / d$ planes
can overlap with each other as well. Since the band touching
energies are different in the two planes, this typically occurs when
there is band touching in one of the planes and almost band touching
in the other one.

\begin{figure}[b]
\centering
\includegraphics[width=8.8cm]{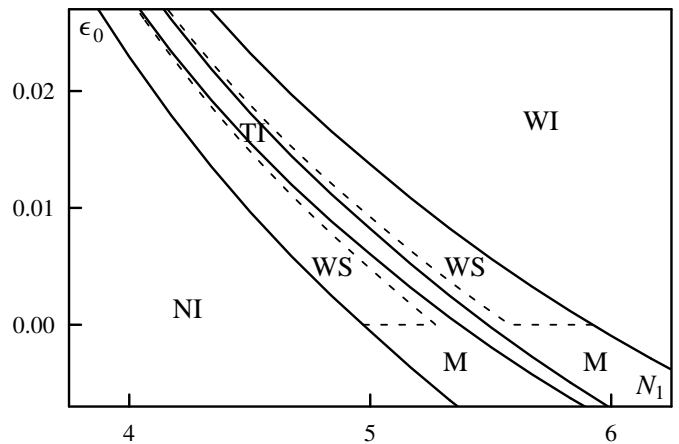}
\caption{Phase diagram of the system against the strain $\epsilon_0$
and the HgTe thickness $N_1$. The other parameters are constant:
$U_0 = 0.2$ eV, $\delta_0 = a / 2$, and $N_2 = 4$. The phase
boundaries separate five distinct phases: the normal insulator (NI),
the strong (3D) topological insulator (TI), the weak (2D)
topological insulator (WI), the band overlap metal (M), and the Weyl
semimetal (WS). The dashed lines indicate approximate phase
boundaries. \label{fig-3}}
\end{figure}

We are now in the position to discuss the other phases around the
Weyl semimetals. The overall transition between the two bulk phases
at small and large values of $\epsilon_0$ is a 2D topological phase
transition because it happens via band touching around both special
points on the $k_z$ axis. This means that the bulk phases on the two
sides of this transition do not have a 3D topological character: we
identify them as the NI phase and the weak (2D) topological
insulator (WI) phase. Since the spin-orbit coupling is stronger in
HgTe than in CdTe, we argue that the system is in the NI phase when
the HgTe layers are thin and in the WI phase when the HgTe layers
are thick.\cite{Bernevig} On the other hand, the phase between the
Weyl semimetals is related to each bulk phase by a topological phase
transition that has a 3D character because it happens via band
touching around only one special point. We conclude that this phase
in the middle is the TI phase. Note that the NI and the WI phases
are equivalent in terms of their 3D topology, therefore there is no
need to distinguish between the two Weyl semimetals: they are
intermediates in two equivalent phase transitions.

The phase diagram of the system against the strain $\epsilon_0$ and
the HgTe thickness $N_1$ is presented in Fig. \ref{fig-3}. Since its
boundaries are interpolated from only 5 points corresponding to
integer values of $N_1$, the phase diagram is only correct on the
qualitative level. Nevertheless, it provides useful guidelines for
the realization of the Weyl semimetal phase in this multilayer
structure. The strain $\epsilon_0$ has to be positive to avoid band
overlap but not too large because that would be hard to achieve
experimentally. This gives a restriction on the thickness of the
HgTe layers: the ideal dimensionless thickness of $4 \leq N_1 \leq
6$ corresponds to an actual thickness of $d_1 \sim 3$ nm, which is
on the border of experimental reasonability.

\subsection{Connection with the superlattice model}
\label{ssec-conn}

To illustrate the relationship with the results obtained in Section
\ref{sec-super}, we discuss the arrangement of the Weyl points in
the Weyl semimetal phase. There are 4 Weyl points that are related
to each other by the symmetries of the system. As $\epsilon_0$ is
gradually increased, and the transition from the NI (TI) phase to
the TI (WI) phase happens through a Weyl semimetal, the Weyl points
first appear at the $k_{x'} = 0$ line, move on approximately
circular curves, and finally disappear at the $k_{y'} = 0$ line.
This is in perfect agreement with the corresponding arrangement for
the superlattice model in Section \ref{ssec-two}. Indeed, the two
models presented in Sections \ref{ssec-two} and \ref{ssec-form} obey
the same symmetries, therefore it is understandable that their
low-energy features are equivalent.

The comparison of the band structures in Sections \ref{ssec-gen} and
\ref{ssec-lin} also makes it possible to estimate reasonable values
for the superlattice parameters in Eq. (\ref{eq-H-1}). The
inversion-breaking voltage $V$ corresponds to the energy difference
$E^{(1)} - E^{(2)}$ between the HOB and the LUB, which is typically
about $0.05$ eV in the tight-binding model. The Fermi velocity of
the surface states becomes $v_F \sim \alpha_{x,y}^{(1,2)} \sim 10^6$
ms$^{-1}$, and the hopping amplitudes are estimated from the typical
energy scale along the $k_z$ axis: $\Delta_{T,N} \sim 10^{-3}$ eV.
The small magnitude of $\Delta_{T,N}$ indicates that the band
structure is relatively flat in the $k_z$ direction. Since $v_r =
v_F \sim 10^6$ ms$^{-1}$, and $d \sim 10$ nm gives $v_z = d
\sqrt{\Delta_T \Delta_N} \sim 10^4$ ms$^{-1}$, this results in the
relation $v_z \ll v_r$ between the effective Fermi velocities around
the Weyl points. Note that the energy scale $\Delta_{T,N}$ also
translates into a maximal temperature $T \sim 10$ K at which the
Weyl semimetal
phase is experimentally observable in this multilayer structure. \\

\section{Physical characteristics} \label{sec-phys}

\subsection{Conductivity anisotropy} \label{ssec-cond}

It was shown in Ref. \onlinecite{Burkov-Balents} that the Weyl
semimetal phase is metallic: when impurities are present, its
conductivity is a finite constant in the limit of zero temperature.
Using the Boltzmann equation, one finds a conductivity $\sigma = e^2
v^2 / 6 \pi \gamma$ for each Weyl point, where $v$ is the effective
Fermi velocity and $\gamma$ is the strength of the impurity
potential. This finite conductivity is a characteristic experimental
feature, especially in contrast with the neighboring NI and TI
phases. In this subsection, we demonstrate that the finite
conductivity at $T \rightarrow 0$ becomes highly anisotropic when
the Weyl semimetal phase occurs due to broken inversion symmetry.

\begin{figure}[h]
\centering
\includegraphics[width=7.4cm]{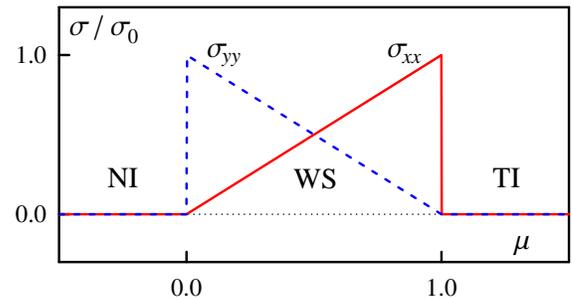}
\caption{(Color online) Variation in the principal conductivities
$\sigma_{xx}$ (red solid line) and $\sigma_{yy}$ (blue dashed line)
during a transition between the normal insulator (NI) and the
topological insulator (TI) phases through the Weyl semimetal phase
(WS). The transition parameter is $\mu = v_F^2 (\Delta_N^{(0)} -
\Delta_T^{(0)}) / [V^2 (\delta_T - \delta_N)]$ and the
conductivities are measured in units of $\sigma_0 = 2e^2 v_F^2 /
3\pi \gamma$. \label{fig-4}}
\end{figure}

To achieve this, we consider the model in Section \ref{ssec-two},
and derive an expression for the conductivity tensor in the limit of
small $\delta_T - \delta_N$. When the condition of the Weyl
semimetal phase is satisfied, there are 4 Weyl points at angles
$\theta_1 = \theta$, $\theta_2 = - \theta$, $\theta_3 = \pi +
\theta$, and $\theta_4 = \pi - \theta$. Due to the convention $0
\leq \theta \leq \pi / 2$ we find that $\theta$ gradually decreases
from $\pi / 2$ to $0$ during a transition from the NI phase to the
TI phase. For each Weyl point labeled by $l$, the conductivity
tensor in the $(x,y,z)$ basis takes the form\cite{Burkov-Balents}
\begin{equation}
\sigma_{l} = \frac{e^2} {6\pi \gamma}
\left( \begin{array}{ccc} v_r^2 \cos^2 \theta_l & v_r^2 \cos \theta_l \sin \theta_l & 0 \\
v_r^2 \cos \theta_l \sin \theta_l & v_r^2 \sin^2 \theta_l & 0 \\
0 & 0 & v_z^2 \end{array} \right), \label{eq-cond-1}
\end{equation}
where we exploit $v_t \ll v_r$ relating the effective Fermi
velocities. Adding the contributions of all 4 Weyl points, there is
a cancelation in the off-diagonal terms, and we obtain
\begin{equation}
\sigma = \sum_{l=1}^4 \sigma_{l} = \frac{2e^2} {3\pi \gamma}
\left( \begin{array}{ccc} v_r^2 \cos^2 \theta & 0 & 0 \\
0 & v_r^2 \sin^2 \theta & 0 \\ 0 & 0 & v_z^2 \end{array} \right).
\label{eq-cond-2}
\end{equation}
As the transition between the NI and the TI phases takes place
through the Weyl semimetal phase, the conductivities in the $x$ and
$y$ directions change in a complementary fashion. In particular,
$\sigma_{xx}$ vanishes on the NI side and $\sigma_{yy}$ vanishes on
the TI side of the Weyl semimetal phase. For an illustration, see
Fig. \ref{fig-4}. The conductivity in the $z$ direction is
approximately constant with $\sigma_{zz} \ll \sigma_{xx},
\sigma_{yy}$ due to $v_z \ll v_r$. Such a strong conductivity
anisotropy that depends sensitively on the system parameters is a
potential hallmark of a Weyl semimetal with broken inversion
symmetry.

\subsection{Topological surface states} \label{ssec-surf}

Since Weyl semimetals are topological phases of matter, they are
characterized by topological surface states.\cite{Burkov-Balents} In
this subsection, we consider the model in Section \ref{ssec-two},
and demonstrate the existence of these surface states. Although we
choose a specific situation and also make a couple of simplifying
assumptions in the following, the topological nature of the surface
states ensures that they exist under more generic circumstances as
well.

In our specific situation, the interface is in the $\{x,z\}$ plane,
therefore any spatial variation is in the $y$ direction only. This
implies that $k_x$ and $k_z$ are still valid quantum numbers. Since
$\Delta_{T,N}$ in Eq. (\ref{eq-delta-2}) do not depend on $k_y$, we
can determine the surface states without taking the explicit
$\mathbf{k}$ dependence into account, and then simply substitute the
appropriate values $\Delta_{T,N}$ for each $k_x$. It is assumed that
only $\Delta_T$ changes with $y$ and the other parameters are
constant: $\Delta_T < \Delta_N$ at $y \rightarrow -\infty$,
$\Delta_T = \Delta_N$ at $y = 0$, and $\Delta_T > \Delta_N$ at $y
\rightarrow +\infty$. Furthermore, if $\Delta_T$ changes
sufficiently slowly, we can approximate it with a linear function in
the important region around $y = 0$: we write $\Delta_T - \Delta_N =
K y$. Expanding the $k_z$ dependent terms up to first order in $k_z'
\equiv k_z - \pi / d$, we find that the surface states $| \Psi
\rangle$ with energy $E$ need to satisfy
\begin{eqnarray}
E | \Psi \rangle &=& \bigg[ v_F \tau^z (-i \sigma^x
\partial_y - \sigma^y k_x) + V \tau^z \nonumber \\
&+& K y \, \tau^x + k_z' d \, \Delta_N \, \tau^y \bigg] | \Psi
\rangle, \label{eq-eigen-1}
\end{eqnarray}
along with $| \Psi \rangle \rightarrow 0$ in the limits of $y
\rightarrow \pm \infty$. To make the subsequent discussion of the
surface states more transparent, we introduce the dimensionless form
\begin{eqnarray}
\tilde{E} | \Psi \rangle &=& \left[ \tau^z (-i \sigma^x
\partial_{\tilde{y}} - \sigma^y \kappa_x) + \tilde{V} \tau^z +
\tilde{y} \, \tau^x + \kappa_z \tau^y \right] | \Psi \rangle,
\nonumber \\ && \label{eq-eigen-2}
\end{eqnarray}
where the variables $\tilde{y} = y \sqrt{K / v_F}$, $\tilde{E} = E /
\Lambda$, $\tilde{V} = V / \Lambda$, $\kappa_x = v_F k_x / \Lambda$,
and $\kappa_z = k_z' d \, \Delta_N / \Lambda$ are all dimensionless,
while $\Lambda = \sqrt{K v_F}$ is a characteristic energy scale.

As a starting point in our discussion, we consider the limit of
$\tilde{V} = 0$. In this case, there are two distinct surface state
solutions for each $\kappa_x$ and $\kappa_z$ that take the analytic
form
\begin{equation}
| \Psi \rangle = \left( i, \mp i e^{i \varphi}, \mp e^{i \varphi}, 1
\right) \, \exp \left( -\frac{\tilde{y}^2}{2} \right)
\label{eq-surf-1}
\end{equation}
in the (T$\uparrow$, T$\downarrow$, B$\uparrow$, B$\downarrow$)
basis, where $\tan \varphi = \kappa_x / \kappa_z$ and the letters
T/B stand for the top/bottom surfaces. The corresponding
dimensionless energies $\tilde{E} = \pm \sqrt{\kappa_x^2 +
\kappa_z^2}$ are indicative of surface states with Dirac dispersion
between NI and TI phases of matter. In the more relevant case of
$\tilde{V} \neq 0$, these analytic solutions only find
straightforward generalizations for $\kappa_x = 0$ when
\begin{equation}
| \Psi \rangle = \left( i, \mp i, \mp 1, 1 \right) \, \exp \left(
-\frac{\tilde{y}^2}{2} \pm i \tilde{V} \tilde{y} \right)
\label{eq-surf-2}
\end{equation}
and the dimensionless energies are $\tilde{E} = \pm \kappa_z$. Note
that these surface states decay in an oscillating fashion at $y
\rightarrow \pm \infty$, and the wave vector $k_y = \pm V / v_F$ of
the oscillations corresponds to the radius of the circle in the $k_z
= \pi / d$ plane along which band touching occurs in Section
\ref{sec-super}.

\begin{figure}[h]
\centering
\includegraphics[width=6.4cm]{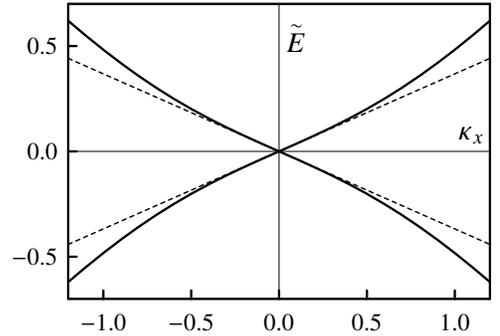}
\caption{Dimensionless energies of the surface states in the
function of the momentum $\kappa_x$ when $\kappa_z = 0$ and
$\tilde{V} = 1$. The dashed lines are linear asymptotes given by Eq.
(\ref{eq-disp-2}) in the $\kappa_x \ll 1$ limit. \label{fig-5}}
\end{figure}

In the most generic case of $\tilde{V} \neq 0$ and $\kappa_x \neq
0$, we solve Eq. (\ref{eq-eigen-2}) numerically and find that there
are still two distinct surface states $| \Psi \rangle$ for each
$\kappa_x$ and $\kappa_z$. The ratios of the vector components in $|
\Psi \rangle$ are no longer independent of $\tilde{y}$, which
explains why simple analytic solutions like those in Eqs.
(\ref{eq-surf-1}) and (\ref{eq-surf-2}) can not be obtained. We
verify that the surface states follow a Dirac dispersion at small
momenta $\kappa_{x,z} \ll 1$, even when the dimensionless voltage
$\tilde{V}$ is large. However, the effective Fermi velocity in the
$\kappa_x$ direction is reduced by a factor that is empirically
found to be $\exp (-\tilde{V}^2)$, and hence the dispersion relation
at small $\kappa_{x,z}$ becomes
\begin{equation}
\tilde{E} = \pm \sqrt{\kappa_x^2 \exp (-2 \tilde{V}^2) +
\kappa_z^2}. \label{eq-disp-2}
\end{equation}
Unlike in the $\kappa_z$ direction where the analytic solution
guarantees the linearity of the dispersion for all $\kappa_z$, there
is a deviation from the linear dispersion in the $\kappa_x$
direction. As illustrated in Fig. \ref{fig-5}, the Dirac dispersion
in Eq. (\ref{eq-disp-2}) is only valid for small enough $\kappa_x$.

If we now restore the $\mathbf{k}$ dependence of $\Delta_{T,N}$, the
surface states $| \Psi \rangle$ remain the same for each $k_x$.
However, they only exist at those $k_x$ for which the difference
$\Delta_T - \Delta_N$ changes sign between $y \rightarrow \pm
\infty$. Assuming without loss of generality that $\delta_T >
\delta_N$ in Eq. (\ref{eq-delta-2}) and that $\Delta_T^{(0)} -
\Delta_N^{(0)}$ does change sign, we find that $\Delta_T > \Delta_N$
for all $k_x$ at $y \rightarrow +\infty$, while $\Delta_T <
\Delta_N$ is only true at $y \rightarrow -\infty$ for $|k_x| < k_0$.
The critical momentum $k_0$ marks the equality $\Delta_T = \Delta_N$
at $y \rightarrow -\infty$, which is one of the conditions required
for band touching in Section \ref{sec-super}. While the material at
$y > 0$ is definitely in the NI phase, the material at $y < 0$ is in
the Weyl semimetal phase if $k_0 < V / v_F$ so that band touching
occurs, and it is in the TI phase if $k_0 > V / v_F$ so that band
touching does not occur. In the former case, surface states exist
between the coordinates $k_x = \pm k_0$ of the Weyl points. As the
Weyl points first appear at the $k_x = 0$ line, and then start to
move further away from it, the range in $k_x$ increases and more
surface states appear. Remarkably, this range characterized by $k_0$
further grows when the material at $y < 0$ is already in the TI
phase, and there is no band touching at all.

\section{Summary} \label{sec-summ}

We proposed a time reversal invariant realization of the Weyl
semimetal phase that occurs due to broken inversion symmetry. We
considered both a superlattice model adapted from Ref.
\onlinecite{Burkov-Balents} and a tight-binding model describing an
experimentally reasonable HgTe/CdTe multilayer structure. The
superlattice model was suitable for analytic calculations due to its
simplicity, while the more realistic tight-binding model required a
numerical treatment.

Although the formulations of the two models are very different,
their identical symmetries lead to equivalent low-energy features.
It should be remarked that, as seen from the generality of the
superlattice model, the Weyl semimetal could be achieved in many
possible material structures.  Exploration of potential compounds
other than HgTe/CdTe would be extremely interesting, especially
given the need to tune strain in the latter to observe the desired
physics. \\

For both models considered, we found a Weyl semimetal phase between
the NI and the TI phases. This phase is characterized by a finite
number of Weyl points with robust band touching at the Fermi level:
the band touching occurs for a finite range of the system
parameters, and hence it can not be removed by an infinitesimal
perturbation. We further verified that the band touching points are
proper Weyl points with a linear dispersion relation around them.

In terms of experimental observation, the potential hallmarks of the
Weyl semimetal phase with broken inversion symmetry are a strong
conductivity anisotropy and the presence of topological surface
states. The highly unconventional low-temperature and low-frequency
bulk transport is discussed in Ref. \onlinecite{BHB}. The Dirac
dispersion relation of the surface states is indicative of TI
materials, but these surface states only exist in a region of
momentum space that is determined by the positions of the Weyl
points. The Weyl semimetal phase between the NI and the TI phases
described in this paper is therefore qualitatively new in terms of
its topological surface states as well. \\ \\

%%%%%%%%%%%%%%%%%%%%%%%%%%%%%%%%%%%%%%%%%%%%%%%%

\begin{acknowledgments}

We are grateful to T. L. Hughes for useful discussions. This
research was supported in part by NSF grants DMR-0804564 and
PHY05-51164 , and by the Army Research Office through MURI grant No.
W911-NF-09-1-0398. G. B. H. acknowledges the support of J. Driscoll
(Trinity College) and the hospitality of KITP during this work.

\end{acknowledgments}

%%%%%%%%%%%%%%%%%%%%%%%%%%%%%%%%%%%%%%%%%%%%%%%%

\end{document}